\documentclass[sn-mathphys,Numbered]{sn-jnl}

\usepackage{graphicx}%
\usepackage{multirow}%
\usepackage{amsmath,amssymb,amsfonts}%
\usepackage{amsthm}%
\usepackage{mathrsfs}%
\usepackage[title]{appendix}%
\usepackage{xcolor}%
\usepackage{textcomp}%
\usepackage{manyfoot}%
\usepackage{booktabs}%
\usepackage{algorithm2e}
\usepackage{listings}%

\begin{document}

\title{Enhancing Stability and Assessing Uncertainty in Community Detection through a Consensus-based Approach}  

\author*[1,2]{\fnm{Fabio} \sur{Morea}}\email{fabio.morea@areasciencepark.it}
\author[2]{\fnm{Domenico} \sur{De Stefano}}
\affil[1]{\orgname{Area Science Park}, \orgaddress{\street{Padriciano, 99}, \city{Trieste}, \country{Italy}}}
\affil[2]{\orgname{University of Trieste}, \orgaddress{\street{Piazzale Europa, 1}, \city{Trieste}, \country{Italy}}}

\abstract{
Complex data in social and natural sciences find effective representation through networks, wherein quantitative and categorical information can be associated with nodes and connecting edges. The internal structure of networks can be explored using unsupervised machine learning methods known as community detection algorithms. The process of community detection is inherently subject to uncertainty as algorithms utilize heuristic approaches and randomised procedures to explore vast solution spaces, resulting in non-deterministic outcomes and variability in detected communities across multiple runs. Moreover, many algorithms are not designed to identify outliers and may fail to take into account that a network is an unordered mathematical entity. The main aim of our work is to address these issues through a consensus-based approach by introducing a new framework called Consensus Community Detection (CCD). Our method can be applied to different community detection algorithms, allowing the quantification of uncertainty for the whole network as well as for each node, and providing three strategies for dealing with outliers: incorporate, highlight, or group. The effectiveness of our approach is evaluated on artificial benchmark networks.}


\maketitle
\section*{Introduction}
Networks provide valuable representations for complex datasets, and community detection serves as a fundamental method to explore their inner structure. A widespread approach to find cohesive groups (henceforth communities or clusters) in social networks aims to identify a densely interconnected subset of nodes compared to the other possible subsets. This principle is quite general although other attachment and aggregation mechanisms are possible in social networks \cite{menardidestefano}. Many methods exist to detect meaningful community structures based on the density of their interconnections \cite{fortunato20yearscd}. The main strategies for this task include the detection of actors or edges with high centrality \cite{NewmanGirvan} optimization-based algorithms \cite{Danon2005}, statistical inference using stochastic block models \cite{lee2019review}, dynamic process-based approaches such as random walks \cite{RosvallPNAS}. Furthermore, a new class of community detection methods has emerged that exploits node semantics or node attributes in addition to network topology. According to the taxonomy proposed by \cite{jin2021survey}, these include graphical model-based community detection, deep learning-based community detection, as well as node embeddings \cite{vskrlj2020embedding}.

Although many of these methods focus on partitioning networks into non-overlapping communities, there is a diverse range of variants, including hierarchical clustering \cite{clauset2008hierarchical}  \cite{plos_Perlasca_hierarchical}, which captures structures at different scales, overlapping communities  \cite{palla2005overlapping} \cite{benatioverlapping} \cite{Ponomarenko_overlapping} and mixed-membership communities \cite{airoldi2008mixed}, where a node can belong to more than one community, as well as a combination of overlapping and non-overlapping communities \cite{moradan2023ucode}. However, probably due to their ability to produce easily interpretable results, optimisation methods that generate non-overlapping partitions are still widely used.

Research has investigated detectability thresholds, resolution limits (which limit the ability to find small communities in large networks), the generation of disconnected communities \cite{traag2019leiden}, and the computation time and cost on large networks.  
In networks with simple topologies, the outcome of different community detection algorithms tends to be consistent. However, networks featuring fuzzy community structures may lead to significant variability among algorithmic results and even within repeated runs of the same algorithm.
Ensemble and Consensus approaches have been suggested with the notion that pooling results from multiple community detection algorithms, or running the same algorithm multiple times, enhances stability and reliability of outcomes.

Given that community detection is an unsupervised machine learning task without a ground-truth, determining which is the 'best solution' based on any parameter is inherently challenging. Consequently, the pursuit of a stable and repeatable result is better sought through algorithmic convergence, based on consensus techniques. In particular, in \cite{lancichinetti2012consensus} the authors, borrowing the idea of consensus clustering, propose its use in the context of complex networks. In their paper, the consensus partition they obtain gets much closer to the actual community structure than the partitions derived from the direct application of a given clustering method. Other approaches to consensus-based clustering for networks can be found in \cite{burgess2016link} or in the literature related to the ensemble clustering algorithms (see, for instance, \cite{ensemble2019}, \cite{evkoski2021community}, and \cite{Evkoski_2021}).
Despite these recent proposals, scientific literature has paid little attention to some issues related to the consistency of community detection results, especially under specific conditions (like increasing fuzziness of the community structure, variability of results in terms of quality of partition, and number of communities). 

In this paper we propose a novel procedure, namely Consensus Community Detection (CCD), that aims to enhance the stability of results of any given community detection algorithm. CCD is based on the assumption that uncertainty is a characteristic feature of the analysis, rather than a flaw in the algorithms and that it is essential to quantify and incorporate it as a valuable part of the results. The procedure starts by generating multiple partitions that differ only due to stochastic factors. A novelty of the approach is that partitions that deviate significantly from the majority are pruned, and the remaining partitions are compared by a consensus procedure based on the co-occurrence of nodes in the same community. Finally, communities are identified as blocks within the adjacency matrix, adding two additional novel features: the identification outliers and the assessment of the uncertainty of the attribution of each node to the final community structure. 

The paper is organised as follows. After an introduction to basic concepts and notations reported in section \ref{basicconcepts}, in section \ref{issues} we discuss the above mentioned issues in the consistency of community detection results and we explore how these issues are relevant for the interpretation of community structure and node membership. In section \ref{ccd} we describe the details of the proposed CCD procedure. Finally, section \ref{results} illustrates its ability to enhance the reliability of results of any community detection algorithm, the use of uncertainty coefficient, and the results obtained on diverse benchmark networks even compared to similar approaches.  

\section{Basic concepts in community detection}
\label{basicconcepts}

\subsection{Notation}
\label{notation}

A network can be modeled by a graph $G = (V, E)$, where $V$ is the set of vertices or nodes ($|V| = n$) and $E = {(i, j) : i, j \in V}$ is the set of edges ($|E| = m$). 
A partition of $G$ is a collection of an arbitrary number of non-empty, pairwise disjoint subnetworks $C$, whose union is exactly $G$ such that
\begin{gather*}
\begin{cases}
  G_1 \cup G_2 \cup \dots \cup G_l \cup \dots\cup G_k = G \\  
 G_i \cap G_j = 0   \forall  i,j \in (1,k)
\end{cases}
\end{gather*}

A community detection algorithm $A(G,\alpha)$ is a function that, given a network $G$, and a set of user-defined parameters $\alpha$, identifies an optimal partition $G$, which best reflects the chosen definition of a community. While a number of different definitions of community exist, in this paper, we assume that a community is a cohesive subset of nodes within a network characterised by a higher degree of internal connectivity compared to external connections, reflecting a tendency for nodes within the community to be more strongly interconnected with each other than with nodes outside the community. It is important to note that, according to this definition, a community is constituted by two or more nodes, hence single-node communities are to be considered either as outliers or as non-valid output. 
Additionally, if the network is composed of two or more separated components, these definitions hold for each individual component.
The output of a community detection algorithm $A$ is a vector of membership labels $C$ that represents an optimal partition of $G$ in $k$ communities:
$$C = A(G,\alpha) = [c_1, c_2, \dots , c_l, \dots, c_n]  \text{   \textit{ membership labels}  } $$ 
The fuzziness of a network partition $C$ can be  measured using the \textit{mixing parameter} $\mu$  defined as:

$$ \mu = \frac{\sum_i{d_i^{ext}}}{\sum_i{d_i^{total}}} $$

where $d_i^{ext}$ is the external degree of node $i$, which corresponds to the number of edges connecting node $i$ to other nodes in different communities, and $d_i^{total}$ is the total degree of node $i$. Consequently, $\mu$ takes values between 0 and 1. The mixing parameter, $\mu$, takes low values in networks with well-defined community structures, where there are minimal connections between different communities. Assuming that a community is defined as a sub-network that exhibits more connections within itself (intra-community) than with the rest of the graph (inter-community) when $\mu$ exceeds 0.5, the partition $C$ identifies communities with more inter-community edges than internal connections, contradicting the definition of community.

Modularity is a quantitative measure that is commonly used in network analysis to evaluate the quality of a partition \cite{newman2006modularity}. Specifically, modularity compares the actual number of edges within a community with the expected number of edges one would observe if the network maintained the same number of nodes, and each node retained its degree, while edges were randomly distributed. Modularity is a suitable measure for evaluating results within the same network, leading to the development of a successful family of community detection algorithms centered on modularity optimization. However, its lack of an absolute interpretation renders it unsuitable for comparing partitions across different networks.

\subsection{Algorithms for community detection} 
Community detection algorithms can be based on a variety of concepts such as optimization of an objective function, similarity metrics, centrality measures, spectral decomposition, random walks, density, deep learning and others. We will test five well-established community detection algorithms: Infomap (IM), Leiden (LD), Louvain (LV), Label Propagation (LP) and Walktrap (WT). These algorithms have been tested and discussed by literature, as in~\cite{Evkoski_2021}, and~\cite{ensemble2019}. 

The Louvain (LV) algorithm~\cite{blondel2008Louvain} optimizes modularity using a greedy approach. Initially, each node is assigned to a separate community; nodes are then iteratively moved to the community of one of their neighbors, maximizing the positive impact on modularity, until no further improvement can be made. LV yields stochastic results, as it relies on random initialization to determine the sequence in which nodes are examined, and identifies a local maximum of modularity. The algorithm has one parameter, called resolution ($r$) that controls the size of detected communities: $r>1$ leads to smaller and more numerous communities, while $r<1$ leads to larger and fewer communities.

The Leiden (LD) algorithm, as introduced in Traag et al.'s work~\cite{traag2019leiden}, is a community detection algorithm primarily designed as an enhancement of the Louvain method, to mitigate the generation of disconnected communities. Notably, it shares similarities with the LV algorithm, employing a resolution parameter and yielding stochastic results.

The Infomap algorithm~\cite{rosvall2007infomap, Rosvall_2009mapequation, mapequation2023software} exploits the information-theoretic duality between finding community structure in networks and minimizing the description length of a random walker's movements on a network; communities are aggregated following an approach similar to LV, using a new random sequential order at each iteration, hence results are stochastic.

Walktrap~\cite{pons2005walktrap} is a hierarchical clustering algorithm based on the assumption that nodes within a  community are likely to be connected by shorter random walks. Beginning with a non-clustered partition, it merges adjacent communities minimizing the squared distances between each node and its community, iterating until no further improvement is possible. A user-defined parameter $s$ defines the length of the random walk to be performed, controlling the resulting community size.

Label Propagation (LP) relies on the notion of proximity or neighborhood relationships, as discussed in~\cite{raghavan2007labelprop}. Initially, each node is assigned a unique community label, then nodes are iterated through in a random sequential order, and each node adopts the label that is most prevalent among its neighbors. This process continues until each node shares the label of the majority of its neighbors. 

\section{Stability and Uncertainty in Community Detection}
\label{issues}
Discovering communities is not an objective per se; rather, it is a tool for interpreting the community structure of the network. The interpretation phase deals with questions such as: How many communities are there in the network? What is the distribution of community sizes - in particular, are there any dominant communities or trivially small ones? Given two nodes in a network, can they be confidently assigned to the same community? Which nodes within the network play distinctive roles, such as leading large and stable communities, or acting as bridges that reduce the distance between otherwise distant or disconnected communities?

We argue that shifting the focus to the interpretation of communities places more stringent demands on the results of community detection. Specifically, given a network $G$ and an algorithm $A$ with user-defined parameters $\alpha$, the resulting partition $C = A(G, \alpha)$ should be (i) consistent with the definition of a community underlying $A$, (ii) should not vary upon repeated executions and (iii) should be insensitive to the specific formulation of the network within a programming environment. 
However, these requirements are often unmet in real-world applications, prompting an examination of four critical issues: the validity and variability of results, the identification of outliers, and the bias introduced by the order of input data. Our investigation will be conducted using the well-known Zachary's Karate network~\cite{zachary1977karate}, and two families of artificial benchmark networks, the Lancichinetti–Fortunato–Radicchi benchmark (LFR) and the so-called Ring of Cliques (RC). 

LFR benchmark networks, proposed by~\cite{lancichinetti2008benchmark} are widely used as benchmarks for testing the performance of community detection algorithms as they are characterised by a power-law distribution of the degree of the nodes (parameter $\tau_1$) and the size of the communities (parameter $\tau_2$).  For the purpose of this paper we use a family of benchmark networks with N = 1000 nodes, with parameters $\tau_1 = 2$, $\tau_2 = 3$, setting an average degree = 10, community size between 20 and 50, and nominal mixing parameter in the range $\mu \in (0.05,0.50)$. Lower values of mixing parameter $\mu$ indicate that the communities are sharply separated and are therefore easily identified by community detection algorithms; on the contrary, high values of $\mu$ are related to networks with fuzzy communities that are hard to identify. 

RC is a benchmark network composed of $k_0$ identical cliques of size $s$, where pairs of cliques are connected in a regular sequence to form a ring. A family of RCs, with a fixed $s$ and varying $k_0$, provides a valuable benchmark for community detection as it ensures a consistent degree of fuzziness with a mixing parameter $\mu  = 1/s!$. A RC is apparently a straightforward problem for community detection algorithms, which can be expected to identify each clique as a community. However, it can become a more challenging problem when additional nodes are introduced such as 'bridge nodes' between pairs of cliques or a central node connected to each clique. Such additional nodes will result in a slight increase in  $\mu$ (while keeping it independent of $k_0$), and create a dilemma for the community detection algorithm since bridge nodes are equally connected to two communities and central nodes are symmetrically connected to each clique. 

All tests are carried out using R language \cite{R-base}, and i-graph library \cite{igraph}. Results are evaluated with Normalized Mutual Information (NMI), a similarity measure between pairs of partitions $NMI(C_1, C_2) \in [0.0,1.0]$, that is computed trough the R-library aricode \cite{NMIaricode}, and the normalized number of communities $k/k_0$. The optimal result for both indicators is $1.0$.

\subsection{Validity of results}
\label{validity}
A community detection algorithm $A$ always returns a partition $C_A$ (i.e. a set of membership labels associated with nodes) but generally it does not provide an explicit assessment of the validity of the result, i.e. whether $C_A$ conforms to the underlying definition of a community as intended by $A$. In the extreme case of a random network, that by definition does not contain any valid community structure, the appropriate result would be to return a single community ($k = 1$). This happens with some algorithms (in our case IM and LP). However, other algorithms as LV, LD, and WT instead produce a $k > 1$ and $\mu > 0.5$. This observation suggests that before interpreting the results of community detection results, we should assess the validity as $(k \ > 1) \land  (\mu \leq 0.5)$.

\subsection{Variability of results}
\label{variability}
Community detection algorithms often use heuristic and randomised approaches to explore large solution spaces, which can lead to different results in successive runs of $A(G, \alpha)$, even when using exactly the same values for $G$ and $\alpha$ are used.
  
\begin{figure}[b!]
\centering
\includegraphics[width=1\linewidth]{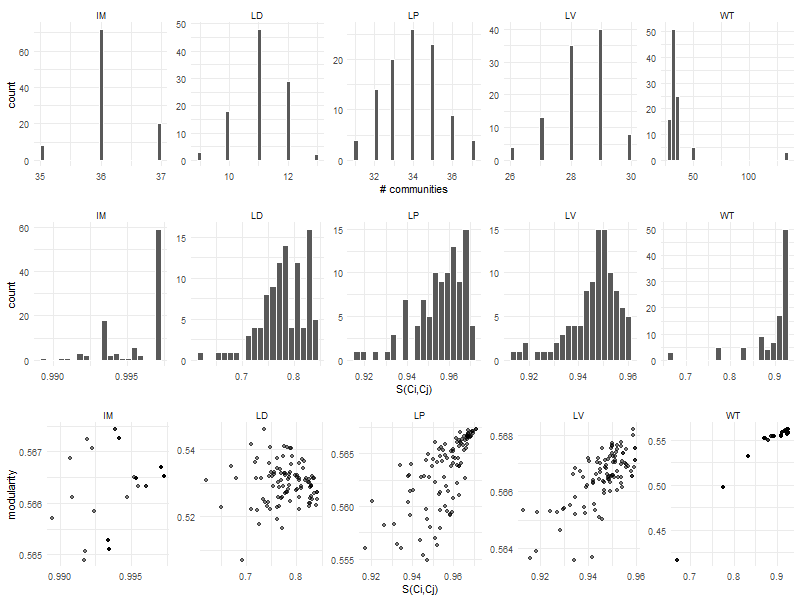}
\caption{\bf{Variability of results of selected community detection algorithms on a LFR benchmark network with a nominal mixing parameter $\mu = 0.40$. Top:  distribution of the number of communities. Middle: similarity between pairs of partitions. Bottom: scatterplot modularity and similarity.}}
\label{fig:1}
\end{figure}

Figure~\ref{fig:1} illustrates the variability of results obtained by different algorithms (LV, LD, IM, WT and LP) on a LFR benchmark network characterised by a nominal value of mixing parameter  $\mu = 0.40$. Partitions and number of communities are different at each trial, and modularity is not sufficient to identify a single optimal solution. Similar results hold for any synthetic or real-world network that exhibits some degree of fuzziness and suggest that relying on a single execution of an algorithm may not consistently and reliably determine the number of communities or assess that any pair of nodes belongs to the same community. 
Variability is more pronounced in large networks with highly interconnected communities, while it may be negligible in simple, small networks. If the objective of the analysis is to answer questions such as "Do nodes $n_1$ and $n_2$ belong to the same community?", variability can pose a significant problem: the answer changes each time we run the algorithm, leading to unreliable and non-reproducible results.  Nevertheless, we argue that variability should not be regarded as a flaw in the algorithm, but as a useful feature that allows deeper insight into the network structure.

\subsection{Outliers}
\label{outliers}
A third observation aims to investigate the behavior of community detection algorithms when facing \textit{outliers}, i.e. nodes that exhibit notably distinct behavior compared to other nodes. Outliers can be highly relevant for interpreting the community structure, for example in a social network it is the case of an individual that is well-connected to many actors that belong to different communities.

\begin{figure}[ht]
    \centering
\includegraphics[width=0.9\linewidth]{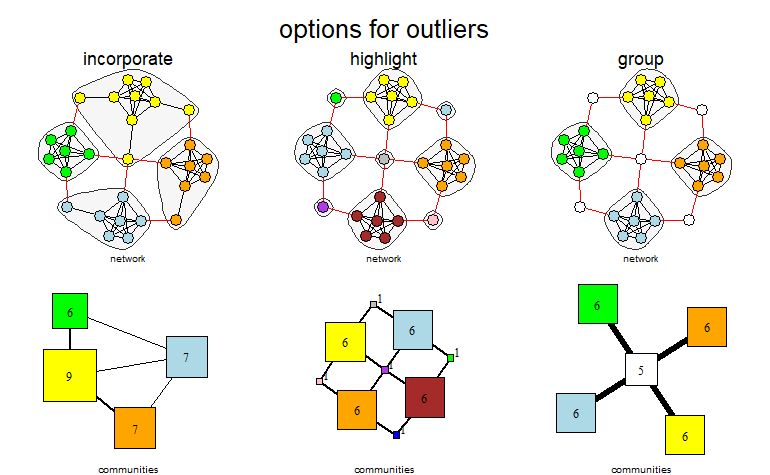}
    \caption{Three alternative strategies to manage outliers: incorporate (left), highlight as single-node communities (center), or group into an outliers' community (right). The top row shows the network; the bottom row shows a graph of the communities, labeled with the number of nodes in each community.}
    \label{fig:2}
\end{figure}

Not all algorithms have the capability to detect outliers. For example, algorithms based on modularity maximization consistently form communities with two or more nodes, and outliers tend to merge into larger communities. Other algorithms have the opposite behavior and place outliers in single-node communities. To categorize these varied approaches, we propose to classify these algorithmic responses to outliers as either "incorporate" or "highlight". Additionally, we introduce a third distinct type of response, termed "group", which involves the identification of individual outliers and their collective assignment to a specific 'community of outliers'.

A RC with $k_0 = 4$, $s = 6$, bridge nodes, and a central node provides a clear representation of the three alternative ways to manage outliers, as shown in Figure \ref{fig:2}. In this case, \textit{incorporating} outliers correctly detects the number of communities ($k=k_0$), but overestimates $s$ and does not capture symmetry. On the other hand, \textit{highlighting} perfectly recognizes community size $s$, but at the cost of overestimating their number ($k = (2 k_0)+1$). Finally, \textit{grouping} provides a trade-off between the previous options, capturing community size and symmetry while adding only a fixed bias to their number ($k=k_0+1$).

\subsection{input-ordering bias}
\label{orderbias}
\begin{figure}[h!]
    \centering
    \includegraphics[width=1\linewidth]{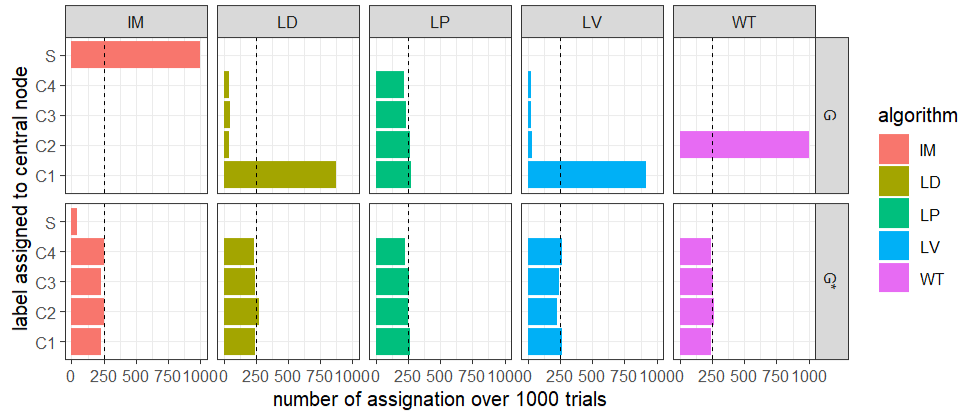}
    \caption{An illustration of input-ordering bias, using a RC with $k_0 = 4$, $s = 5$ with bridges and a central node. Above: label assigned to the central node by various algorithms, applied $t=1000$ times to  network $G$. Below: labels assigned to the central node applied to network $G^*$, a copy of $G$ randomly permuted at each iteration. Labels: S = the center is highlighted as a single-node community, $Ci$ = the center is incorporated in community $i$.}
    \label{fig:3}
\end{figure}

Networks are inherently non-ordered, but their practical representation in a computer model is inevitably ordered. Ideally, community detection algorithms should ignore order, but this is not always the case in practice. The issue can be highlighted by comparing $C = A(G)$ (i.e. the results produced by a community detection algorithm $A$ to network $G$) with $C^* = A(G^*)$, where ${G^*}$ is generated by a random permutation of edges and vertices of $G$. If $A$ is unbiased algorithm, we may expected that $C = C^*$. In complex, real-world networks the differences $C$ and $C^*$ may not be noticed. However, the bias can be devised using a network with nodes and edges built in sequential order, a sharply defined and symmetric community structure, and identifiable outliers, such as the RC depicted in Fig~\ref{fig:2}. The central node is connected with equal strength to four communities, hence one would expect that an unbiased algorithm assigns it to any of the four communities with equal likelihood. Fig~\ref{fig:3} shows the results of a test on different algorithms, over 1000 iterations.  

We observe that most algorithms exhibit a noticeable input-ordering bias, with the exception of LP. Specifically, when applied to $G$, IM assigns the center to a single-node community, WT always to the same community $C_2$, while LD and LV strongly favor community $C_1$. However, when applied to $G^{*}$, each algorithm produces a less biased result. This example highlights the influence of the network's inherent ordering on outcomes, emphasizing the need to use $G^{*}$ to reduce bias. Input ordering bias has been discussed in the literature, notably by \cite{OrderNature, OrderGood, EnsembleDetection} mainly focusing on modularity-based methods. In this paper, we aim to generalize these results to any algorithm, and to devise a procedure that mitigates input-ordering bias, while improving the stability and reliability of results.

\section{Consensus Community Detection}
\label{ccd}
The conventional approach to community detection entails selecting an algorithm and executing it to derive a vector of membership labels $C^A$, which is then interpreted as the optimal partition of $G$ into communities. 
If repeated executions of $A$ lead to different outcomes $C^A_1 \neq C^A_2 \ldots \neq C^A_t  $, a single "optimal partition" can be identified as the one that maximizes an objective function $C^A_{opt} = \arg\max_{C_i} M(C_i^A)$. A good candidate for $M$ if modularity, which emerges naturally if $A$ is a modularity-based algorithm such as LV or LD. This approach yields a single solution, but not a stable one, as it may be surpassed by subsequent iterations of the procedure. Furthermore, there exists no clear correlation between modularity optima and other pertinent features, such as the number of communities, as illustrated in Figure \ref{fig:1} (bottom row). 

Consensus offers a more robust option to enhance stability. For example, as discussed in  \cite{lancichinetti2012consensus}, distinct partitions obtained by repeated execution of $A$ are used to build a co-occurrence matrix, $D$, in which each entry $d_{ij}$ signifies the proportion of partitions in which vertices $i$ and $j$ are clustered together. $D$ is then interpreted as an adjacency matrix for a new network, representing the community structure. In the new network, edges below a chosen threshold $p$ are pruned, and the process is repeated recursively until $D$ is a block-diagonal matrix, where each block is interpreted as a community. The process is effective but has the disadvantage of requiring multiple iterations. Moreover, pruning can generate disconnected nodes (vertices that have all edges below the threshold $p$), hence a threshold $p = 0.6$ is recommended. To maintain network connectivity, disconnected nodes are aggregated into the neighbouring community with the highest weight.  The algorithm's ability to identify communities of varying scales and its capacity to properly identify outliers is limited by this assumption, as discussed in section \ref{outliers}.

Another approach can be found in ~\cite{evkoski2021community} and \cite{Evkoski_2021}, which propose the Ensamble Louvain algorithm to find stable communities. As with the previous method, a co-occurrence matrix $D$ is calculated, and communities are identified by pruning with a threshold $p = 0.9$. Selecting such a high threshold value returns more stable results without necessitating recursive iterations but has the drawback of overlooking outliers, i.e. all the nodes that fall above the threshold. Depending on the network topology and the objective of the analysis, outliers may be a negligible minority. However, we argue that they should be considered appropriately, as discussed in section \ref{outliers}. 

Other consensus approaches have been presented, to address specific issues. For example,~\cite{burgess2016link} is focused on incomplete networks, and leverages a link-prediction strategy to infer missing intra-community edges and casting results with a consensus approach. Ensemble methods involve combining the outcomes of multiple community detection algorithms. One notable example is the ensemble method introduced by~\cite{EnsembleDetection} that aims to identify overlapping and fuzzy communities.
 
Our research introduces a novel Consensus Community Detection (CCD) procedure, that can be applied to any community detection algorithm, to produce a stable representation of communities, improving the reliability and interpretability of results.  Specifically, CCD addresses the four challenges outlined in section \ref{issues}, dealing with the validity of detected communities, reducing variability across different algorithm runs, quantifying the residual variability, dealing with outliers, and mitigating the input-ordering bias.

While the variability of clustering results is widely explored in data science, its direct application to network community detection has been less emphasised in the literature. Notably, the specific considerations regarding the handling of outliers and input-ordering bias in the context of community detection have been largely overlooked.

CCD is based on the assumption that uncertainty is an inherent characteristic of community detection, hence it should be carefully assessed and incorporated into the results. For this reason, CCD represents the community structure in the form of a matrix $\widetilde{C}$ of size $2 \times n$, that associates each node with a community label and an uncertainty coefficient:
 
\begin{gather*}
\widetilde{C} = 
\begin{cases}
[c_1, c_2, \dots , c_l, \dots, c_k]  \text{   \textit{ membership labels}  }\\
[\gamma_1, \gamma_2, \dots , \gamma_l, \dots, \gamma_k]  \text{ \textit{  uncertainty coefficients}  } 
\end{cases}
\end{gather*}

where $\gamma \in [0,1]$ represents the degree of uncertainty associated with the assignation of a membership label to each node. $\gamma = 0$ indicates that the corresponding node is always co-occurring in the same community of at least one other vertex of the community. Higher values of $\gamma$ indicate that the vertex was associated with different communities at each trial of the community detection. 

Our approach builds on previous work but differs in three major aspects: (1) addresses input-ordering bias, (2) introduces the novel \textit{uncertainty coefficient} $\gamma$ serving as a concise representation of residual variability at the node level, which can be subsequently leveraged for in-depth network analysis, and (3) introduces a quantile threshold $q$ to select the partitions according to a similarity score, which allows for faster and more stable computation.

CCD provides a comprehensive framework to augment the efficacy of existing community detection algorithms, hence it maintains compatibility with legacy methods, enabling straightforward comparisons with prior analyses and established literature.

At a high level, the CCD procedure can be delineated into three overarching steps: partition generation, pruning, and consensus. 

During the first step, $t$ independent partitions of $G$ are calculated, as $C_i = A(G^*,\alpha)$ where  $G^*$, a randomly permuted version of $G$. If a partition is valid (i.e. $k > 0.5$ and $\mu <= 0.5$), it is appended to the list of $L$. At the end of the loop, if $L$ is empty, the algorithm returns a null partition, signifying that under the given algorithm $A(\alpha)$ there are no valid solutions. Otherwise, in the second step, the algorithm computes the similarity score for each partition $S_i=mean_j(NMI(C_i,C_j))$ for each partition in $L$ and prunes the list by removing partitions with similarity scores below a predefined quantile threshold $q$. In the third step, the algorithm proceeds with the assignment of community labels to nodes, along with uncertainty coefficients. This step operates iteratively until all nodes have been evaluated: within each iteration, a block $B \in D$ is identified based on a predetermined threshold $p$ within the co-occurrence matrix $D$. Nodes within this block are assigned a community label and an uncertainty coefficient $\gamma$ computed as the average value of $b_{ij}$ within the block. The final result is a partition $\widetilde{C^A}$ providing a representation of the community structure along with associated uncertainty measures.

The pseudocode for our procedure is represented in Algorithm \ref{alg:ccd}.

\RestyleAlgo{ruled}
\SetKwComment{Comment}{/* }{ */} 
\begin{algorithm} 
\caption{Consensus Community Detection}\label{alg:ccd}
\KwData{Network $G$, Algorithm $A(\alpha)$, Parameters $p$, $q$, $t$}
\KwResult{Partition $\widetilde{C^A}$ }
\For{$i = 1 $ to $t$}{ 
    $G^* \gets$\text{permute}$(G)$ \Comment{ shuffle the network}
    $C \gets A( G^*, \alpha )$  \Comment{ apply algorithm $A$ to obtain partition}
    $\mu \gets \text{mixing parameter of }C$ \;
    $k \gets \text{number of communities in } C$\;  
    \If{$(k > 1) \land (\mu > 0.5)$}{
         Add $C$ to the list of partitions $L$\;
    }
}
$n_p\gets \text{number of partitions in} L$\;
<\eIf{$n_p > 0$}{
    \For {each partition $C_l$ in $L$}{
        $S_l = \text{mean}(NMI(C_l, C_m))$ for $m \in (1,n_p)$ \Comment{Similarity score}
    }
    Prune $L$ removing partitions with $S$ below quantile threshold $q$\;
    $D \gets 0$ \Comment{initialize co-occurrence matrix} 
    \For{each partition $C_l$ in $L$}{
        \For{each community in the partition}{
            \For{each pair of nodes $i$, $j$ in the community}{
                \If {nodes $i$, $j$ are assigned to community $k$}{
                $d_{ij} \gets d_{ij} + 1$
                } 
            }
        }
    } 
    $D \gets D / n_p$ \Comment{normalize co-occurrence matrix } 
    $c \gets 0$ \Comment{initialize community labels}
    $k \gets 0$ \Comment{initialize community counter}
    \While{any nodes to be evaluated}{
        $k \gets k + 1$\;
        Identify block $B \subset D$ such that $b_{ij} \leq p $\;
        \For{each node $j$ within block $B$}{
            Assign community labels $c_j \gets k$\;
            Assign uncertainty coefficient $ \gamma_j = 1 - \text{mean}(b_j)$\;
        }   
    }
    $\widetilde{C^A}  \gets  [c, \gamma]$ \Comment{community labels and uncertainty coefficients}
}{
    $\widetilde{C^A}  \gets \text{NULL}$  
}
\end{algorithm}

\section{Results}
\label{results}
In this section, we show the results of tests on CCD to address all the issues shown in \ref{issues}, namely its ability to reduce the variability of results, assess uncertainty, identify outliers, and reduce the input-ordering bias. Finally, we evaluate the performance of CCD in identifying a known community structure in three cases: (1) Karate network; (2) a family of RC networks with a fixed $\mu$, but varying numbers of communities, and (3) a family of LFR networks with varying value of the mixing parameter $\mu$.

\subsection{Reduction of variability - parameter t}
\label{results_variability}
CCD operates through a repetitive process executed for a designated number of iterations, denoted as $t$. Our first test concentrates on evaluating residual variability in relation to $t$, a critical decision involving a trade-off between cost (increasing linearly with $t$) and performance. We utilize two metrics: the count of identified communities ($k$) and the similarity between all pairs of partitions, assessed with NMI.

\begin{figure}[h]
\centering
\includegraphics[width=1.0\linewidth]{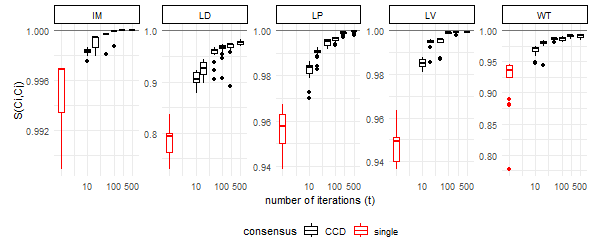}
\caption{Stability of CCD results as a function of the number of iterations $t = (10, 20, 50, 100, 200, 500)$. Results of single trials $t = 1$ are highlighted in red. Test on a LFR network with $\mu = 0.3$, CCD parameters $p = 0.8$ and $q = 0.5$. Stability is measured by the similarity between pairs of solutions  $S(C_i,C_j) = mean(NMI(C_i,C_j))$.}
\label{fig:4}       
\vspace{-0.4cm}
\end{figure}

The test is conducted on LFR benchmark networks with parameters as outlined in section \ref{issues} and a nominal mixing parameter of $\mu = 30$. CCD was implemented with $p = 0.6$, $q = 0.5$, and $t$ values ranging from 5 to 500. Stability, measured as the similarity between pairs of partitions, ideally yields $S = 1.0$. Results, depicted in Figure~\ref{fig:4}, reveal that CCD significantly enhances stability compared to single trials, with stability increasing as $t$ increases, gradually reducing dispersion and approaching the optimal value. Notably, each algorithm reaches a plateau at a distinct value of $t$. In practical applications, the choice of an optimal $t$ involves a trade-off between result stability and computational resources, where the right balance depends on the interplay between the network characteristics, the chosen algorithm and the analysis objectives.

\subsection{Assessment of residual variability}
\label{results_uncertainty}
To illustrate how CCD assesses uncertainty associated with nodes, we apply it to the Karate network mentioned in section \ref{notation}.  We use the LV algorithm with $t = 100$, and different values of the resolution parameter $r$ to control the granularity of community structure.
\begin{figure}[h]
\centering
\includegraphics[width=.89\linewidth]{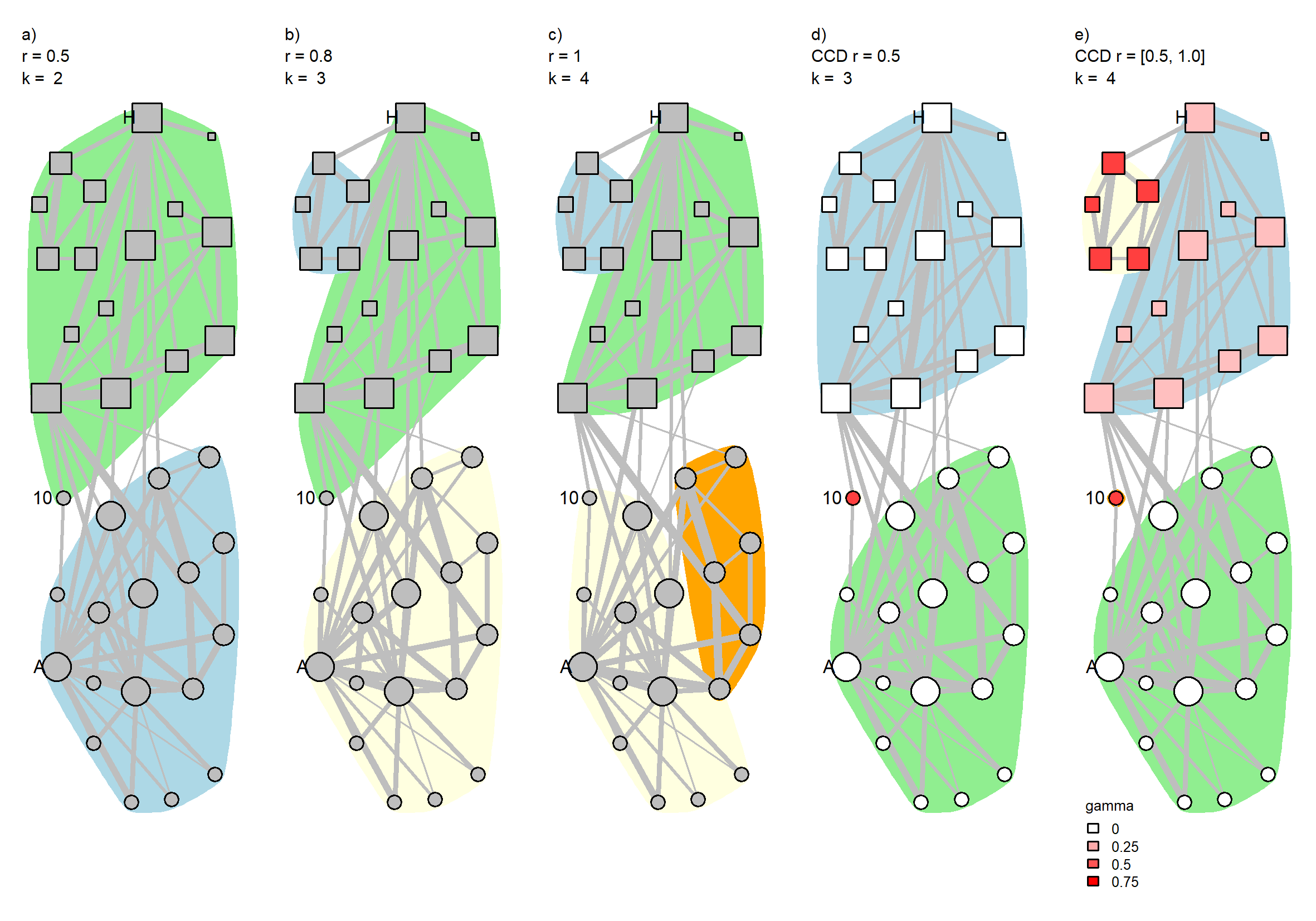}
\caption{Example of CCD Zachary's Karate network (weighted). a) single trial of Louvain with resolution $r = 0.5$. b) single trial of LV, $r = 0.8$. c) single trial of LV, $r = 1.0$. d) CCD with $t = 100$ and $r = 0.5$ e) CCD with $t = 100$ and $r \in [0.5, 1.0]$. Uncertainty coefficient $\gamma$ is available only for CCD.}
\label{fig:5}       
\end{figure}

Results are shown in Figure \ref{fig:5}, where some nodes are labelled: H and A (leaders of the two main communities) and node 10 (that may belong to either community, depending on the chosen value of $r$ and the random variation that characterizes each trial). The first three panels display the results of single trials, demonstrating how the number of communities $k$ depends on the resolution parameter $r$. For example, in panel a) with $r = 0.5$ the result is $k = 2$; in panel b) with $r = 0.8$ there are two distinct results: $k = 3$ (in $61\%$ of trials), and $k = 2$ ($39\%$ of trials). As per panel c), setting resolution $r = 1.0$ leads to $k = 4$. In the context of unsupervised machine learning, all the above results are equally valid. However, even when the value of the $r$ is fixed, there is still significant variability that hinders interpretation.   Panel d) shows how CCD can improve the interpretability: selecting $r = 0.5$, $p = 0.9$ and $q = 0.5$, produces a simple community structure with $k = 2$ and highlights node 10 as an outlier, with an uncertainty $\gamma = 0.75$, expressed by the color scale. Panel d) showcases a more nuanced application of CCD, where the resolution parameter assumes a different value at each trial, randomly selected in the range $[0.5, 1.0]$ which allows identification of $k = 3$ communities at different scales, and associates different levels of uncertainty to each.

\begin{figure}[h]
    \centering
    \includegraphics[width=.9\linewidth]{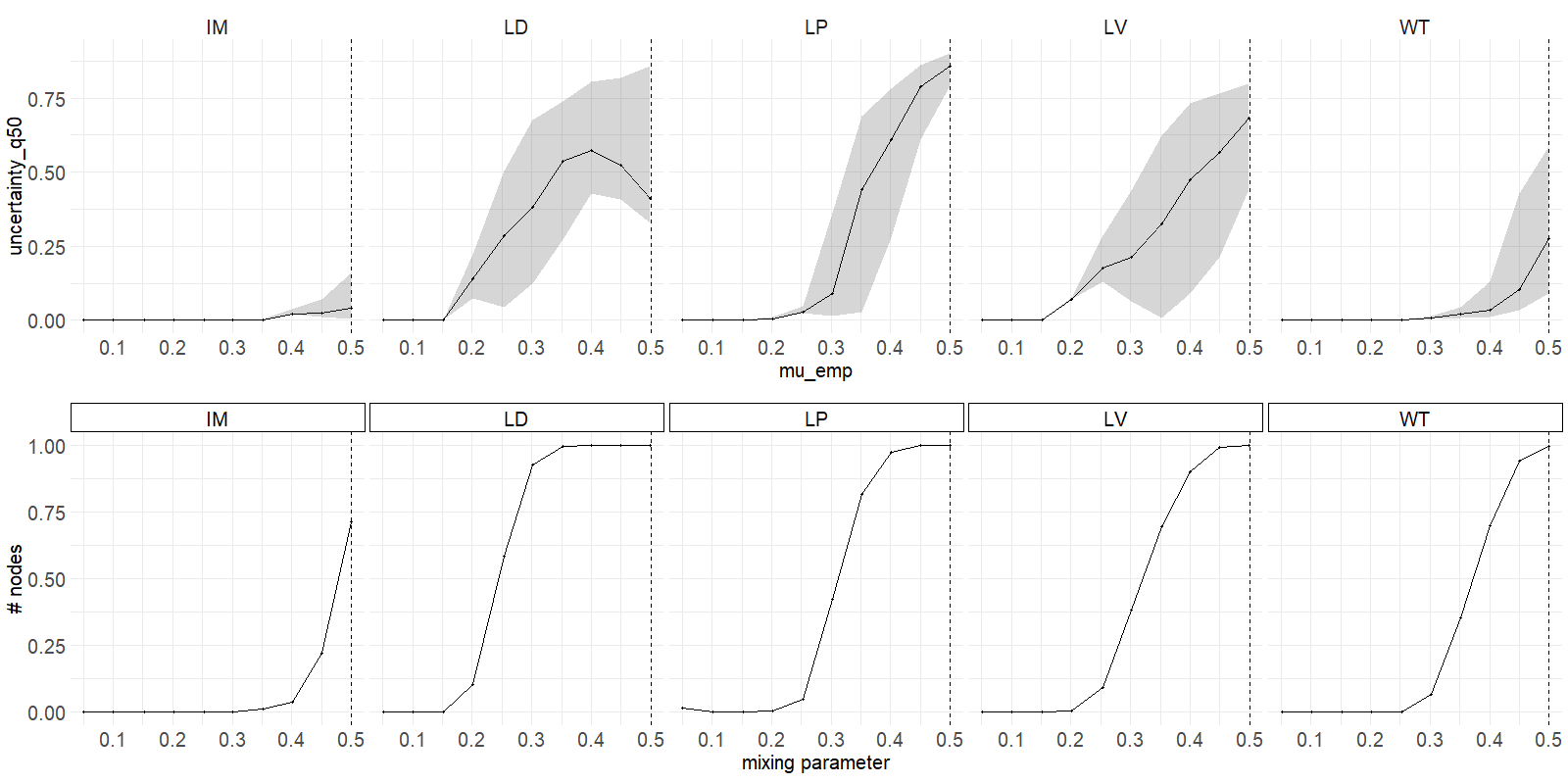}
    \caption{Assessment of uncertainty with CCD on a family of LFR benchmark networks. Top: median value of $\gamma$; the shaded area is delimited by $10^{th}$ and $90^{th}$ percentile. Bottom: fraction of nodes with $\gamma > 0$. }
    \label{fig:6}
\end{figure}

Uncertainty is assigned at each node, but it can be summarised at the network level by the number of nodes with some degree of uncertainty or by the mean value of the uncertainty coefficient. Figure~\ref{fig:6}) shows both measures for a set of LFR with the characteristics presented in section \ref{issues}; specifically the top row shows the fraction of nodes with $\gamma > 0$ and bottom row the median value of $\gamma$; the shaded area is delimited by $10^{th}$ and $90^{th}$ percentile. In both cases uncertainty increases non-linearly with the mixing parameter. However the behavior is different according to the algorithm: in this example, IM identifies the communities with almost no variations for $\mu < 0.4$, then increases sharply. Other algorithms show a growing number of uncertain assignations at low values of $\mu$, and plateau for $\mu > 0.3$.  
  
\begin{figure}[h] 
   \centering
     \includegraphics[width=.9\linewidth]{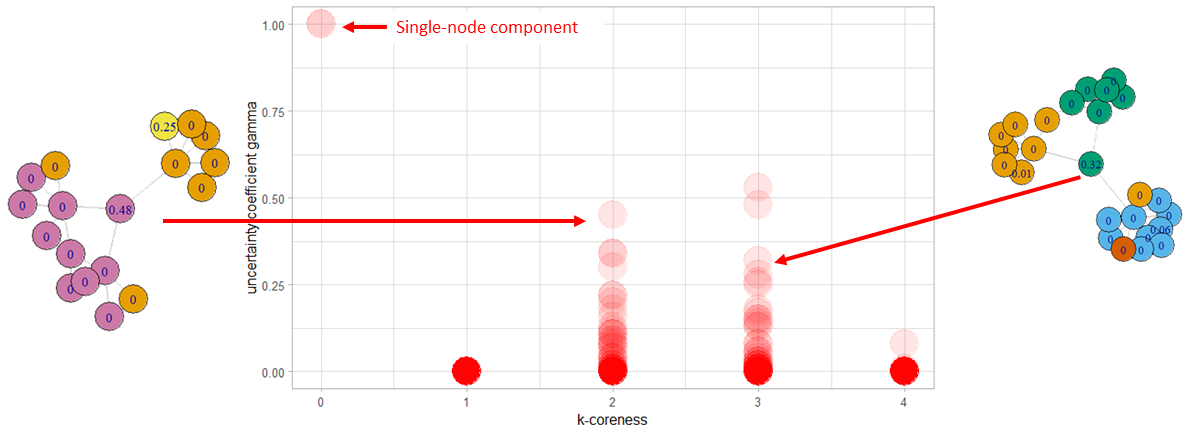}
     \caption{A coreness-uncertainty diagram on LFR network with $\mu = 0.40$. The subgraphs on the left and right of the diagram show the neighborhood of two nodes with high uncertainty.}
     \label{fig:7}
 \end{figure}

Figure~\ref{fig:7} shows a possible use of $\gamma$ in unraveling network structure, in conjunction with a centrality centrality measures - in this example  \textit{k-coreness}, a centrality measure introduced by \cite{kong}. Specifically, a \textit{k-core} is a subgraph where all vertices are connected to at least \textit{k} other vertices within that subgraph; the \textit{k-coreness} of a node indicates the highest k-core that the node belongs to. 
 The example is calculated on a LFR benchmark network with $\mu = 0.4$, and communities are detected CCD with parameters $t = 1000$,  $p = 0.6$, and $q = 0.5$.  
 The scatterplot depicts \textit{k-coreness} against $\gamma$;  two examples of nodes with high uncertainty are highlighted by the arrows, and their respective neighborhood (at geodesic distance equal to 2) is depicted as subgraph. A single-node component is represented in the top left corner of the scatterplot($k-coreness = 0$ and $\gamma = 1$).  

\begin{figure}[ht]
    \centering
    \includegraphics[width=1\linewidth]{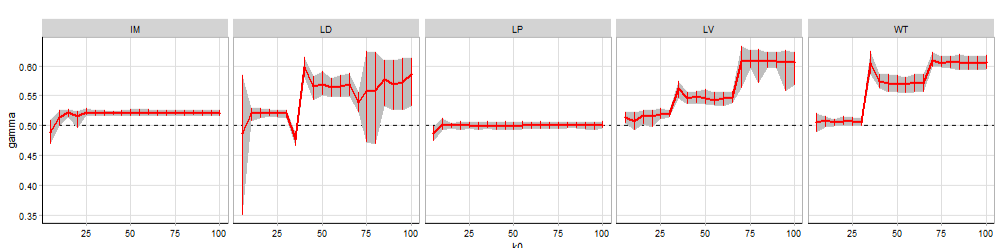}
    \caption{Uncertainty coefficient assigned by CCD to the bridge nodes of a family of RC benchmark networks. The expected value $\mu = 0.5$ is highlighted by the horizontal dotted line.}
    \label{fig:8}
\end{figure}

Finally, a specific test is carried out to assess the ability of different algorithms to assign an appropriate value of $\gamma$. To ensure a reproducible example with a known expected value, the test is conducted on a family of RCs with clique sizes $s = 6$ and a number of cliques in the range $k_0 \in \{5, \ldots, 100\} $. CCD was applied using $p = 0.8$, $q = 0.5$ and $t = 200$.  The test is focused on the bridge nodes, i.e. nodes that connect two successive cliques within the ring, and can be expected to have  $\gamma = 0.5$.   Fig~\ref{fig:8} shows that most algorithms behave well within a limited range of $k_0$, and that for larger rings there are remarkable variations depending on the algorithm. IM and LP produce very stable results even for $k_0 = 100$.   

\subsection{Assessment of performance}
\label{perf}
In this section we evaluate the performance of CCD in identifying a known community structure, focusing on the ability to determine the number of communities and measuring the similarity between the inherent community structure and the outcomes of community detection. 

The first test evaluates the ability of CCD to detect communities of varying sizes, on a family of LFR benchmark networks with parameters presented in section \ref{issues},  Performance is assessed with two indicators: NMI (similarity between the identified communities and the built-in communities), and the normalized number of communities ($k/k_0$). CCD parameters are $t = 1000$, $q = 0.5$

\begin{figure}[t]
    \centering    \includegraphics[width=.9\linewidth]{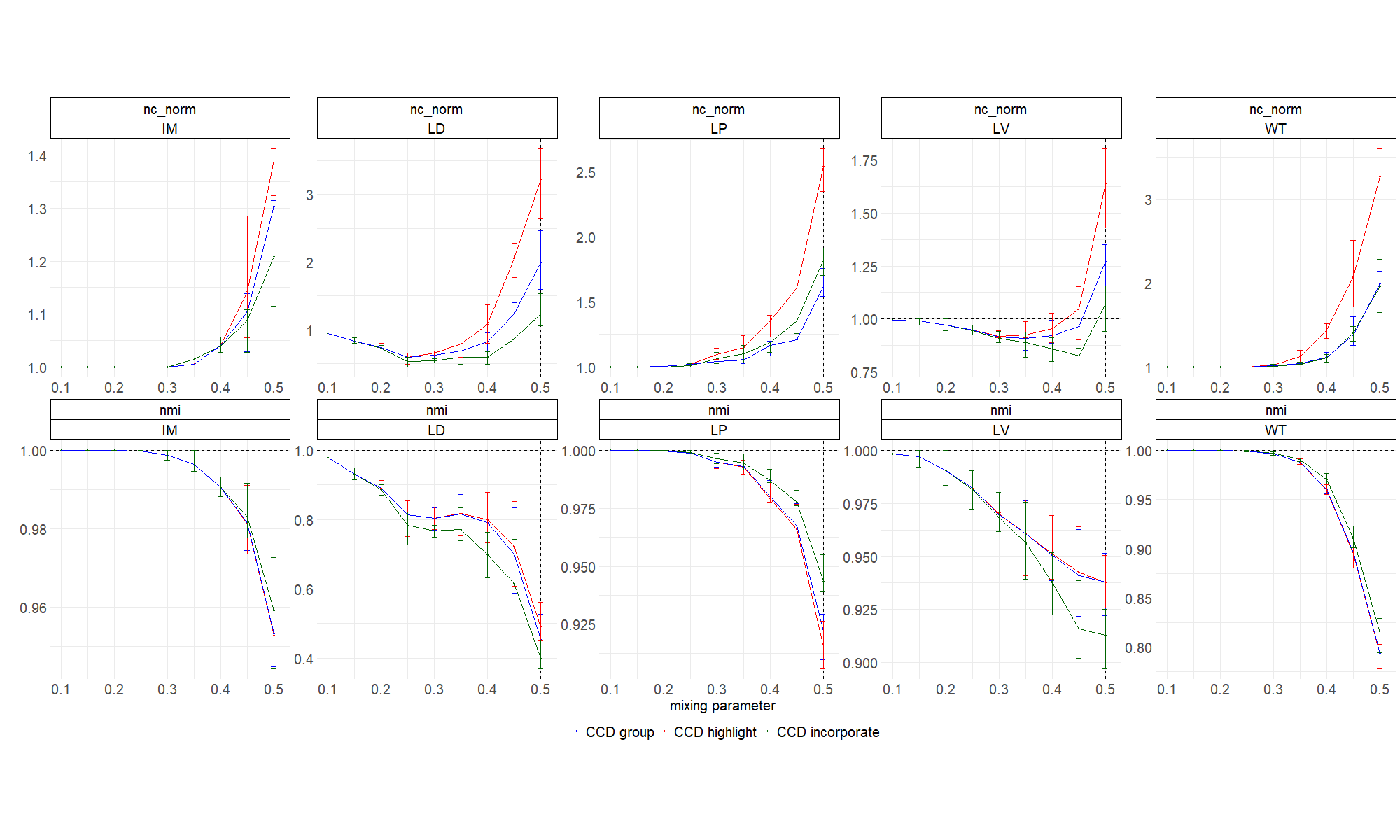 }
    \caption{Performance of CCD  on a family of LFR benchmark networks, using different strategies to manage outliers: group (blue), highlight (red) or incorporate (green).}
    \label{fig:9}
\end{figure}

Figure \ref{fig:9} compares the performance of the three different strategies to manage outliers discussed in section \ref{outliers}: for low $\mu$, the curves overlap, indicating no significant deviations; however, as $\mu$ increases, differences emerge.  Incorporating outliers ($p=0.6$) leads to the best performance in terms of  $k/k_0$, but may hinder performance measured by NMI, especially with modularity-based methods LV and LD. On the other hand, highlighting ($p=0.8$) generates several single-node communities, resulting in lower performance in terms of $k/k_0$, but may offer an advantage in the interpretation of results. Finally, grouping ($p=0.8$ and all outliers re-assigned to community $0$) provides a trade-off between the previous options:  it captures community structure (NMI comparable to previous case), still allows for the identification of outliers and adds smaller errors to $k/k_0$.

\begin{figure}[ht!]
    \centering    \includegraphics[width=.9\linewidth]{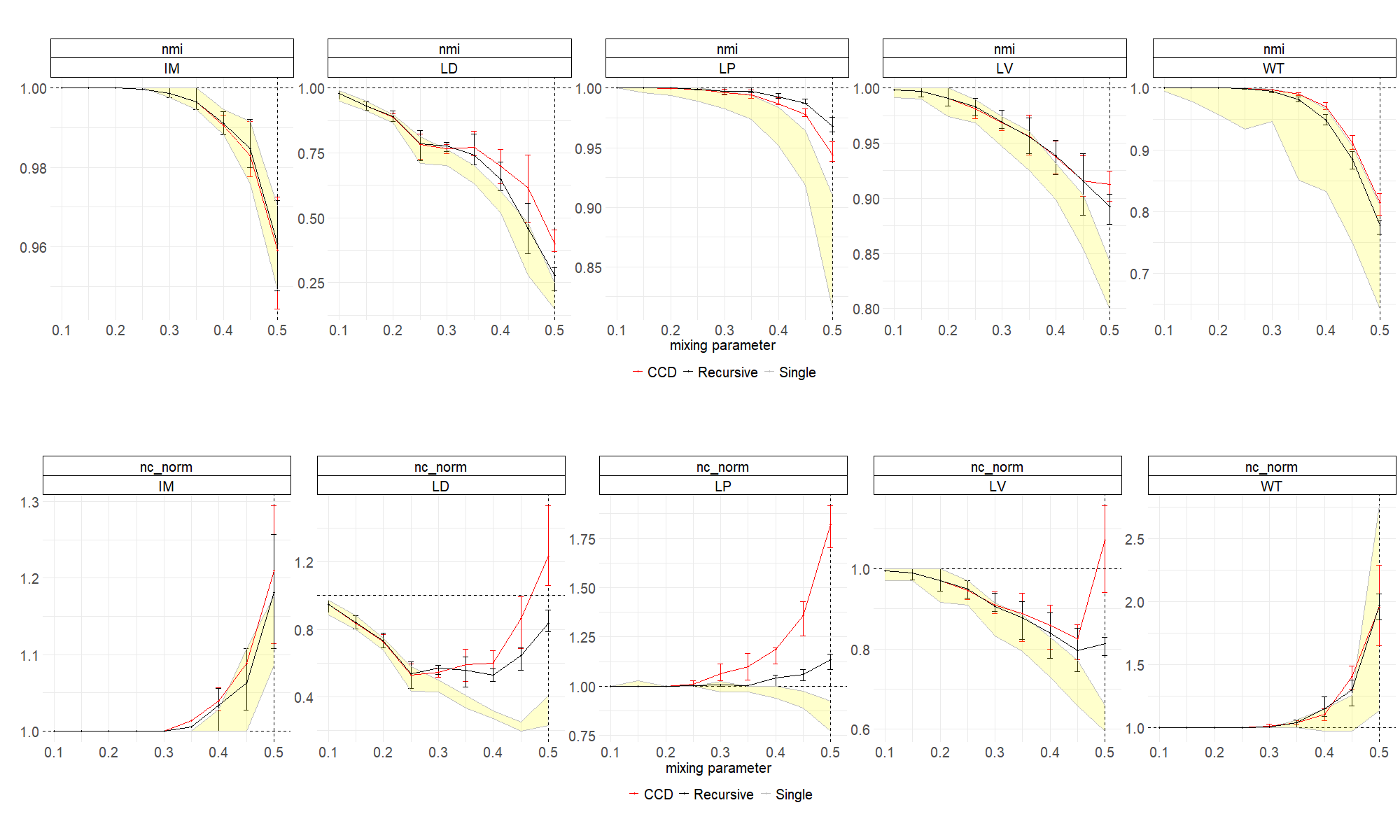 }
    \caption{Performance of CCD on a family of LFR benchmark network. CCD (red), is compared to recursive consensus (black) and single trials (yellow).}
    \label{fig:10}
\end{figure}

Figure \ref{fig:10} compares the performance of CCD (incorporating outliers, $p = 0.6$) with single trials and the recursive consensus community detection technique introduced by Lancichinetti et al.~\cite{lancichinetti2012consensus}.  When measuring performance with NMI, consensus methods are outperforming single trials, especially as $\mu$ increases, although with different behavior depending on the algorithm. Performance measured with $k/k_0$ is comparable for WT and IM and diverges for the other methods as the fuzziness of the benchmark network approaches the limit value of $\mu = 0.5$.

The second test is focused on the effectiveness of identifying small, non-overlapping communities of the same size. The test is performed on a family of RC where $k_0$ varies between 5 and 100; CCD parameters are $p = 0.8$, $q = 0.5$, the network is shuffled and outliers are grouped according as discussed in \ref{outliers}.

\begin{figure}[hb]
    \centering    \includegraphics[width=.99\linewidth]{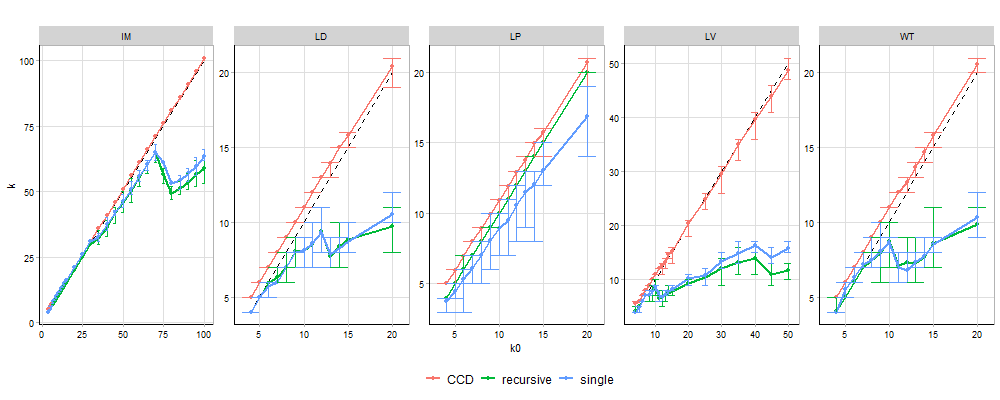}
    \caption{CCD results on a family of RC, compared to recursive consensus and single trials.}
    \label{fig:11}
\end{figure}

  Results are shown in Figure \ref{fig:11}, representing the number of cliques $k_0$ versus the number of communities detected by CCD (red), recursive consensus (green), and single trials (blue). A dashed line shows the ideal result $ k = k_0$. For low values of $k_0$ all methods perform well: the number of communities identified by the algorithm $k$ is equal (or very close) to the number of cliques in the network $k_0$. However, as $k_0$ increases, most algorithms tend to agglomerate neighboring communities, resulting in $k < k_0$, with behavior depending on the algorithm. In all cases, CCD is more accurate than recursive consensus and single trials, even for small cliques $s = 3$ arranged in large rings up to $k_0 = 100$.  In addition, CCD generally provides more stable results, as indicated by the vertical error bars in each plot, and allows the identification of outliers and quantitative assessment of uncertainty.

\section{Conclusions}
\label{conclusions}
In this study, we introduce a novel Consensus Community Detection procedure (CCD) that can enhance the stability of any community detection algorithm and facilitate the interpretation of results. CCD exploits the inherent variability of community detection across individual trials of a given algorithm to derive a metric, denoted \textit{uncertainty coefficient} $\gamma$, which quantifies the uncertainty associated with each node.

Tests were carried out on artificial benchmark networks, which provide an ideal setting for comparing the effectiveness of detecting a known community structure. Our findings indicate that the CCD outperforms single trials in terms of result repeatability, stability, and prevention of invalid and inconsistent results. 

However, in real-life applications, community detection is often employed in unsupervised machine learning scenarios, where there is no predefined structure to be detected or used as a benchmark to measure performance. Here, CCD offers distinct advantages over other consensus procedures, as it enhances stability, facilitates the assessment of residual uncertainty, and provides strategies for managing outliers through aggregation, highlighting, or grouping.  

Potential directions for further research encompass a thorough evaluation of CCD on diverse real-world networks, to determine its adaptability and efficacy across varied contexts. A further area of research is the testing of CCD with community detection methods based on neural networks and deep learning.

\section*{Data and code availability}
The code and benchmark networks used in this research are available at:  https://github.com/fabio-morea/CCD. 

\section*{Acknowledgments}
This research was funded by the European Union - Next generation EU, with the support of the Italian Ministry of University and Research under the PRIN 2022 project 2022MSL3AY "Methods for the analysis of scientific collaboration networks".
 
\bibliography{references} 

\end{document}